\documentclass[a4paper]{spie}  

\usepackage{multirow}
 
\usepackage{amsmath,amsfonts,amssymb}
\usepackage{astro_bib_macro_mam}
\usepackage{graphicx}
\usepackage[colorlinks=true, allcolors=blue]{hyperref}

\title{ANDES, the high-resolution spectrograph of the ELT: simulated performance of the CORO module and overview of the high-contrast capabilities for exoplanet observations}

\author[a]{M. N’Diaye}
\author[b]{A. Simonnin}
\author[a]{A. Spang}
\author[a]{P. Berio}
\author[a]{A. Chiavassa}
\author[c]{G. Agapito}
\author[a]{C. Bailet}
\author[c]{G. Carlà}
\author[a]{Y. Caujolle}
\author[d]{O. Carrión-Gonzalez}
\author[d,a]{G. Chauvin}
\author[e]{S. Cuevas}
\author[f]{O. Gabella}
\author[g]{M. Houllé}
\author[a]{S. Lagarde}
\author[a]{P. Martinez}
\author[c]{E. Pinna}
\author[d,a]{B. Rajpoot}
\author[a]{J. Seidel}
\author[h]{C. Selmi}
\author[i]{A. Vigan}
\author[j]{P. Di Marcantonio}

\affil[a]{Université Côte d'Azur, Observatoire de la Côte d'Azur, CNRS, Laboratoire Lagrange, Nice, France}
\affil[b]{Lund Observatory, Division of Astrophysics, Department of Physics, Lund University, Lund, Sweden}
\affil[c]{INAF - Osservatorio Astrofisico di Arcetri, Firenze, Italy}
\affil[d]{Max-Planck-Institut für Astronomie, Heidelberg, Germany}
\affil[e]{Instituto de Astronomía, Univ. Nacional Autónoma de México, Ciudad de México, Mexico}
\affil[f]{LUPM, Université de Montpellier, CNRS, Montpellier, France}
\affil[g]{Université Grenoble Alpes, CNRS, IPAG, Grenoble, France}
\affil[h]{INAF - Osservatorio Astrofisico di Brera, Merate, Italy}
\affil[i]{Aix Marseille Université, CNRS, CNES, LAM, Marseille, France}
\affil[j]{INAF - Osservatorio Astronomico di Trieste, Trieste, Italy}

\authorinfo{Contact: mamadou.ndiaye@oca.eu}

\begin{document} 
\maketitle

\begin{abstract}
We present the simulated performance of the coronagraph (CORO) module for ANDES, the Extremely Large Telescope (ELT) high-resolution spectrograph. ANDES aims to address a broad range of science cases, including the characterization of the atmosphere of exoplanets. With a first light envisioned by 2035, the instrument baseline features a modular fiber-fed echelle spectrograph with visible and near-infrared ultra-stable spectral arms to provide a simultaneous spectral range of 0.4-1.8\,$\mu$m with a spectral resolving power up to 100,000. ANDES also includes an Integral Field Unit (IFU) mode-fed by a single-conjugate adaptive optics (SCAO) module and an insertable CORO module, enabling the combination of high-contrast imaging and high-dispersion spectroscopy (R=100,000) for the study of exoplanet properties. In this contribution, the ANDES CORO design, its main features and its simulated performance are detailed in the presence of SCAO residual errors to probe exoplanet atmospheres with the spectro-imager mode at high-spatial resolution and with unprecedented angular resolution. We show the preliminary simulation results on the predicted contrast and the expected detection performance with ANDES performance unfolded (APU), the simulation tool to determine the yield of exoplanets that will be detected in emitted and possibly reflected light. The characterization operation of the instrument will be discussed to assess the ANDES ability to detect atomic and molecular signatures connected to the exoplanet atmosphere characteristics.
  
\end{abstract}


\section{INTRODUCTION}
\label{sec:intro}  
ANDES \cite{Marconi2024,DiMarcantonio2024} is an Extremely Large Telescope (ELT) instrument which is designed with several subsystems to provide high-resolution spectroscopy over a broad spectral range extending from the visible to the near-infrared. With an expected first light in 2034-2035, the instrument will achieve a spectral resolution of approximately R=100,000, enabling detailed characterization for a wide range of astrophysical objects such as stellar atmospheres, circumstellar environments, and exoplanets. In particular, to observe exoplanets in emitted or possibly reflected light with direct imaging spectroscopy, the instrument relies on two main upstream methods: (i) adaptive optics to correct for the effects of atmospheric turbulence on the image of an observed star, and (ii) coronagraphy to block the starlight diffracted by the telescope aperture.

ANDES features two arms to feed the spectrographs in different wavelength ranges: a seeing-limited arm and a diffraction-limited arm. The latter is obtained by means of adaptive optics capabilities through the Single Conjugated Adaptive Optics (SCAO) subsystem\cite{Pinna2024} to feed the YJH and the K Integral Field Unit (IFU) spectrographs. To expand the ANDES scientific capabilities, a coronagraph module was officially integrated as part of the SCAO subsystem into the instrument baseline configuration at the ANDES consortium meeting in April 2025. The objective of this module is to enable high-contrast observations combined with high-resolution spectroscopy over the 1.0-1.7$\mu$m spectral range (YJH bands), particularly for the direct characterization of exoplanets at small angular separations\cite{Palle2025}.

The requirement for the coronagraph module is set to provide images with a contrast level of at least $3\times 10^{-3}$ (goal $10^{-3}$) at 20 mas over the YJH bands. Alternatively, very high-resolution spectroscopy techniques, such as the molecular mapping method \cite{Hoeijmakers2018}, can theoretically enable an additional contrast gain from $10^{-3}$ to $10^{-4}$. Combining high-contrast imaging with high-resolution spectroscopy could theoretically enable up to $10^{-7}$–$10^{-8}$ contrast levels\cite{Snellen2015}, which could prove suitable for detecting low-luminosity rocky companions and observing the population of temperate giant exoplanets around nearby stars.

In this contribution, we present the study of the coronagraph design, the optimization process, and its achieved contrast performance in median observing conditions (seeing of 0.67"). A preliminary error budget is then derived to determine the specifications for different sources of errors in order to ensure a stable $10^{-3}$ contrast during the observation. The opto-mechanical design of the coronagraph module is presented in a companion paper by Berio et al.\cite{berio2026} in these proceedings. Finally, we present some preliminary end-to-end simulations of the SCAO+IFU combination with the ANDES Performance Unfolded (APU) simulator (Simonnin et al. in prep) to assess the contrast performance of the instrument and derive some first detection limits for gas giant planets observed in emitted light.

\section{Coronagraph design for ANDES SCAO-IFU mode}
\subsection{Starlight suppression system with a classical Lyot coronagraph}

Assuming the image of an observed on-axis unresolved star, the contrast requirement is set to $3\times 10^{-3}$ (goal $10^{-3}$) at an angular separation of 20 mas from the host star in 1.0-1.7\,$\mu$m wavelength range (YJH bands) with ELT/ANDES in SCAO mode. In the following, the contrast is intended to be computed on the radial profile of the intensity of the star image normalized to its peak and averaged over a radial bin of 3.5\,mas (first one centered on the peak). The latter corresponds to the size of a IFU spaxel of 7mas.

Based on the simulations from the SCAO group using the SPECULA adaptive optics simulation framework\cite{Rossi2026}, the system is expected to deliver images with a Strehl ratio of (initially 77\% at 500\,Hz) 83\% at 1\,kHz in median condition in H-band and using an AO reference star of magnitude larger than 8 in I-band, leading a contrast of $1.2 \times 10^{-2}$ at 20 mas. To gain about one order of magnitude in contrast and reach the required $3\times 10^{-3}$  contrast, one solution consists in adding a starlight suppression system downstream the SCAO system, with a moderate gain in starlight attenuation. Based on these assumptions, the expected amount of residual wavefront errors, and the wide wavelength range of work, we decide to work with a simple coronagraphic system by considering a classical Lyot coronagraph (CLC) \cite{Lyot1932,Vilas1987} in our study. Alternative coronagraph concepts such as vortex phase masks\cite{Mawet2005} could be investigated before the final design review of ANDES if there is a favorable evolution of the initial assumptions with a clear and robust benefit in terms of science return and without additional system complexity.

\begin{figure}[!ht]
    \centering
    \includegraphics[width=\linewidth]{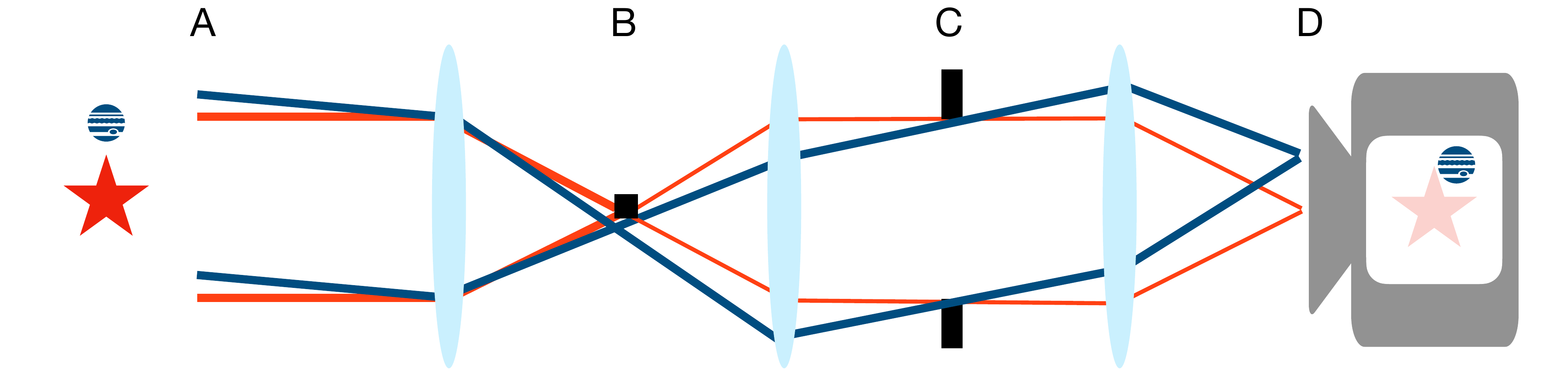}
    \caption{Optical layout of the CLC with four planes: the entrance pupil plane A, the downstream focal plane B in which an opaque mask (FPM) is located, the relayed pupil plane C with a diaphragm (PPM) called Lyot stop, and the re-imaged focal plane D in which the image of the scene is formed.}
    \label{fig:corono_layout}
\end{figure}

The scheme of the CLC is represented in Figure \ref{fig:corono_layout}. We briefly recall its principle, assuming a perfectly conjugated system and no aberrations. Our astrophysical scene includes an unresolved on-axis bright star and its faint planetary companion. The incoming light from both objects is captured by the telescope pupil in plane A which forms the image at the telescope focal plane B. At this location, an opaque focal plane mask (FPM) is set up to block the central part of the star image while leaving the contribution from the planet light nearly intact. In the re-imaged pupil plane C, the residual starlight left is then filtered by a pupil plane stop (PPM) known as Lyot stop to remove the diffraction effects due to the FPM. Finally, the final image of the astrophysical scene is formed onto the camera in the re-imaged focal plane D. An image of the star with attenuated signal is observed, leading to an enhanced contrast that favors the observation of the faint companion.

\subsection{Optimization of the CLC}
The goal is to design a CLC which achieves a $3\times 10^{-3}$ contrast (goal $10^{-3}$) at 20\,mas from the star over the spectral bands of interest (YJH) and offer an appropriate throughput to preserve the photons coming from the planetary companion. In the context of SCAO-ANDES, we need to dimension the two elements of the device, which are the opaque FPM and the Lyot stop.

\begin{figure}[!ht]
    \centering
    \includegraphics[width=0.49\linewidth]{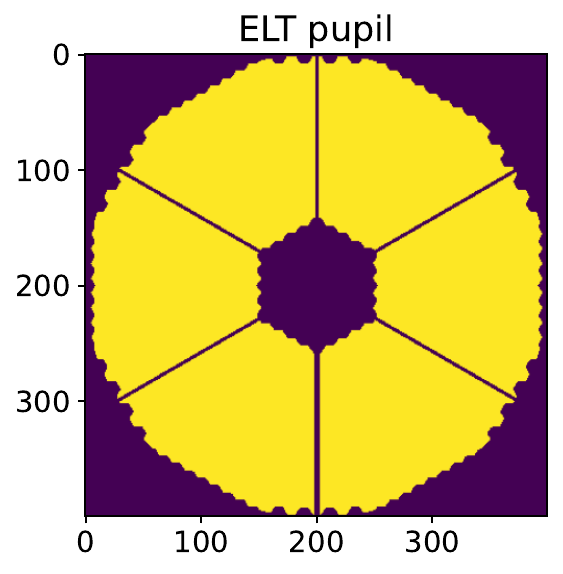}
    \includegraphics[width=0.49\linewidth]{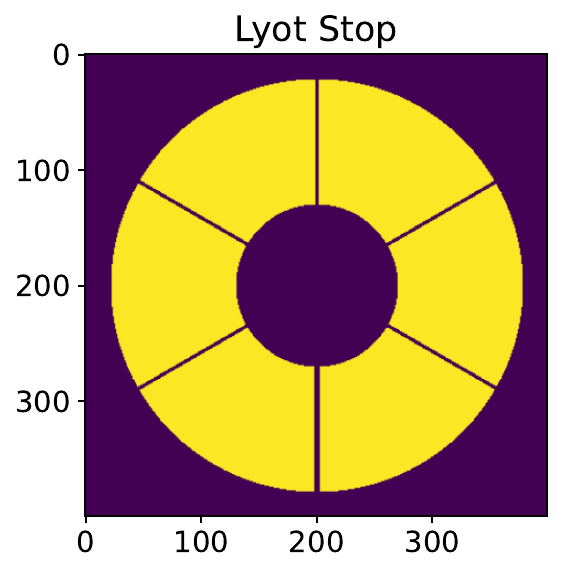}
    \caption{Display of pupils for the CLC. Left: ELT pupil. Right: Example of Lyot stop (final version for the coronagraph).}
    \label{fig:pupils}
\end{figure}

In our present study, the ELT aperture and Lyot stop displayed in Figure \ref{fig:pupils} represent the entrance and exit pupils of our coronagraph. In the following, the term $D$ denotes the diameter of the circumscribed circle of the ELT aperture. The design of the Lyot stop is an annular pupil mask with inner and outer diameters $ID$ and $OD$. It also includes struts which share the same shape as the ELT pupil as an initial assumption. We ran some simulations about the impact of the strut thickness on the contrast and found out that it has limited impact on the contrast yielded by the coronagraph.

To optimize the contrast performance provided by our coronagraph, we perform a parameter space exploration on three parameters: the FPM size $d$, the Lyot stop $OD$ and $ID$. The terms $ID$ and $OD$ are expressed in fraction of $D$.

The coronagraph will act as a spatial filter to remove the light from an observed star. Assuming an unresolved source through a good imaging system, the star image or point-spread function (PSF) presents a time-evolving structure which can be decomposed into two main contributions at first order \cite{Sivaramakrishnan2001,Perrin2003,Aime2004}: (i) a deterministic part that is associated with the diffraction effects due to the telescope aperture and static aberrations, (ii) a random part that is related to the different sources of dynamic aberrations (e.g. atmospheric residuals, instrument optical surface errors, instrumental errors due to thermal or mechanical drifts).

The coronagraph mostly reduces the deterministic part of the PSF that is due to the aperture diffraction effects in the star image. The coronagraph is therefore optimized in contrast and throughput in the absence of aberrations. Its performance can then be evaluated in the presence of aberrations, e.g., under different median seeing conditions. Following this approach and to design the FPM and Lyot stop, we performed a parameter space exploration for these parts to maximize the contrast at 20\,mas in the wavelength range of 1.0-1.7\,$\mu$m and in the absence of aberrations. The contrast is evaluated at 20\,mas over a photometric aperture with a diameter corresponding to the size of a single IFU spaxel, i.e., 7\,mas.

The overall throughput of the instrument represents a key aspect to determine the amount of photons from a planetary companion around a nearby star and assess the throughput budget that can be allocated to the coronagraph. Performing end-to-end simulations of the full SCAO-IFU instrument is a promising solution to estimate the throughput budget for the coronagraph module. Starting in 2022, Simonnin et al. (in prep) have been developing ANDES performances unfolded (APU), a post-AO end-to-end simulation tool to perform contrast estimates and exoplanet yields. A first public release of the code should be made available in 2026 (Simonnin et al. in prep). At this stage and for the sake of simplicity, we look for optimizing the coronagraph throughput with a parameter space exploration of $d$, $OD$, and $ID$ to maximize the signal-to-noise ratio (SNR) between the residual light of an observed star and the light of a planetary companion over a spaxel at 20\,mas from the optical axis.

\subsection{Parameter space exploration}
Our goal now is to determine the optimal parameters for $d$, $ID$, and $OD$. To this aim, we first define a search area in the coronagraphic image of an observed star at an angular separation of 20$\pm$3.5 mas at which the light contribution from the star is assessed. To consider the planet light loss due to the Lyot Stop, we compute the coronagraph throughput by computing the fraction of light of the setup with no FPM, in the absence and in the presence of the Lyot stop. Our simulations are performed in broadband light over the spectral band of interest. 

The parameter space exploration is performed with the following ranges for $d$, $OD$, and $ID$:
\begin{itemize}
\item d ranges between 3.0 and 4.5\,$\lambda_{ref}/D$ with a step of 0.1\,$\lambda_{ref}/D$.
\item OD ranges between 0.82 and 0.96\,$D$ with a step of 0.01\,$D$
\item ID ranges between 0.30 and 0.45\,$D$ with a step of 0.01\,$D$
\end{itemize}
in which, the reference wavelength $\lambda_{ref}$ is set to 1.6\,$\mu$m. Two metrics are considered for the optimization: (i) the contrast which represents the averaged residual starlight intensity $I_s$ through the coronagraph over the 1.0-1.7\,$\mu$m wavelength range, (ii) the inverse of the signal-to-noise ratio SNR$^{-1}$ for an off-axis companion which can be expressed as the ratio between the residual starlight $\eta_s$ and the square of the light of an off-axis companion $\eta_p$ at 20\,mas, i.e. $\eta_s$ / $\eta_p^2$, following the prescriptions given in the literature \cite{Ruane2018b}.

\begin{figure}[!ht]
    \centering
    \includegraphics[width=0.325\linewidth]{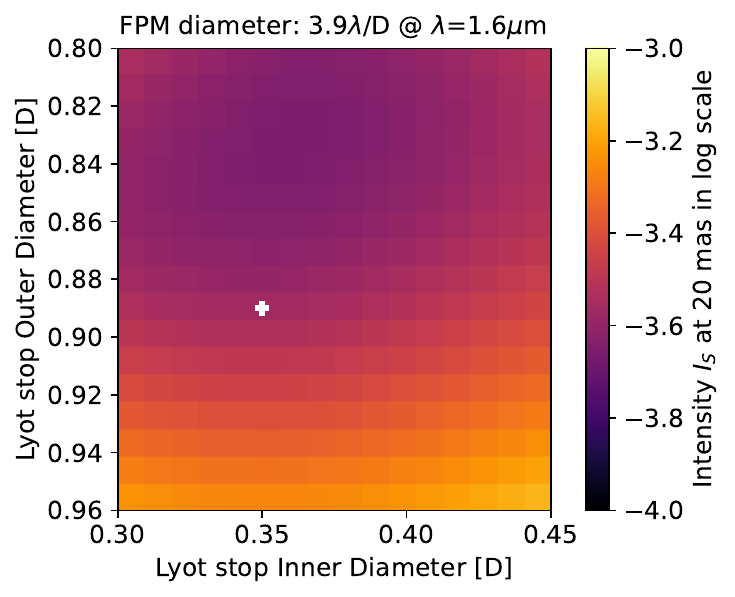}
    \includegraphics[width=0.325\linewidth]{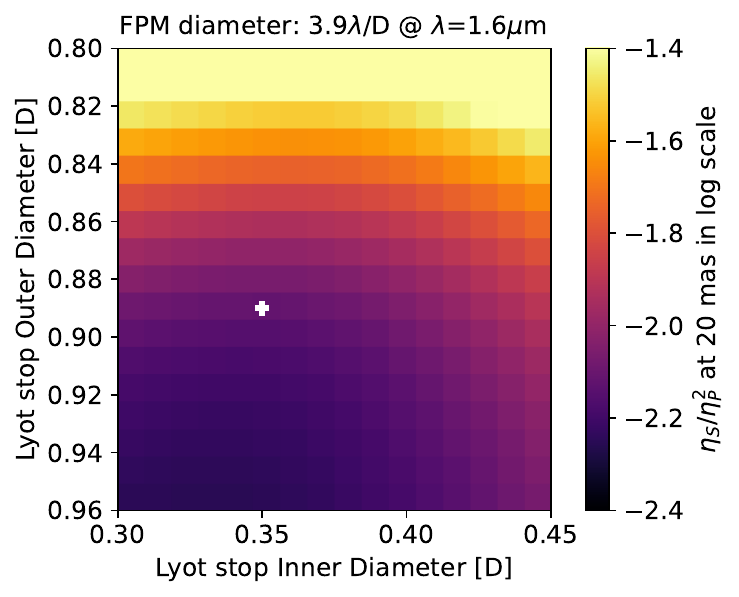}
    \includegraphics[width=0.325\linewidth]{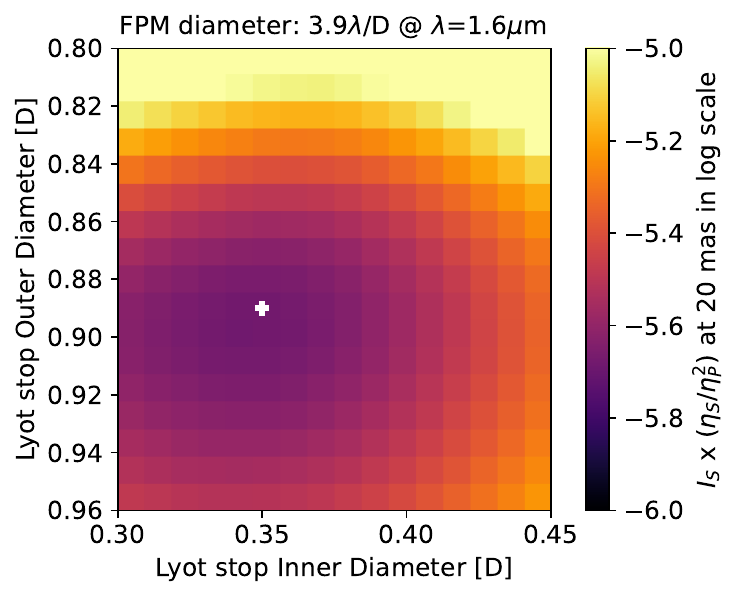}
    \caption{Residual intensity $I_s$ in log scale in broadband light (optimization in YJH band) (left), the SNR$^{-1}$ (middle), and the joint metric (right), inside the coronagraphic image within the search area, i.e., 20.0$\pm$3.5 mas from the star, as a function of the Lyot stop $OD$ and $ID$ for a FPM diameter $d$=3.9\,$\lambda_{ref}/D$ (left). The white cross-mark denotes the selected solution that is recalled in Table \ref{tab:CLC_parameters}.}
    \label{fig:parameter_space_exploration}  
\end{figure}

Figure \ref{fig:parameter_space_exploration} represents the contrast seen as the residual star intensity $I_s$ in the presence of the coronagraph (left panel), the SNR$^{-1}$ seen as the ratio (middle panel) and the joint metric (right panel) as a function of Lyot stop $OD$ and $ID$ for a FPM diameter $d$=3.9$\lambda_{ref}/D$.

The white cross in the three panels represents the optimal solution with the joint metric that was found with the parameter space exploration for the coronagraph design. In the left and middle panel, this solution appears slightly sub-optimal for the residual star intensity $I_s$ and for the SNR$^{-1}$. However, the solution still provides an average contrast of $10^{-3}$ over 1.0-1.7\,$\mu$m and the SNR$^{-1}$ is reduced by a factor smaller than 2 with respect to the optimal SNR$^{-1}$ value in this panel. In addition, the combined joint metric shows that our solution represents an optimal trade-off with both reducing the starlight intensity residuals below $10^{-3}$ and maximizing the SNR for an off-axis companion at 20\,mas from the star. 

Based on this analysis, we found an optimal configuration for the 1.0-1.7\,$\mu$m wavelength range with $OD = 0.89\,D$, $ID = 0.35\,D$, and $d=3.9\,\lambda_{ref}/D$. Table \ref{tab:CLC_parameters} summarizes these results for YJH configuration.

\begin{table}[!ht]
    \centering
    \caption{Parameters of the CLC for the YJH band configuration.}
    \begin{tabular}{c|c|c|c}\hline\hline
        \multirow{2}{4em}{Wavelength range} & FPM & \multicolumn{2}{|c}{Lyot Stop in $D$} \\
         & Diameter $d$ in $\lambda_{ref}/D$ and mas & Outer diameter $OD$ & Inner diameter $ID$\\\hline
         1.0-1.7\,$\mu$m (YJH bands) & 3.9 and 33.4 & 0.89 & 0.35\\\hline
    \end{tabular}
    \label{tab:CLC_parameters}
\end{table}

\begin{figure}[!ht]
    \centering
    \includegraphics[width=0.67\linewidth]{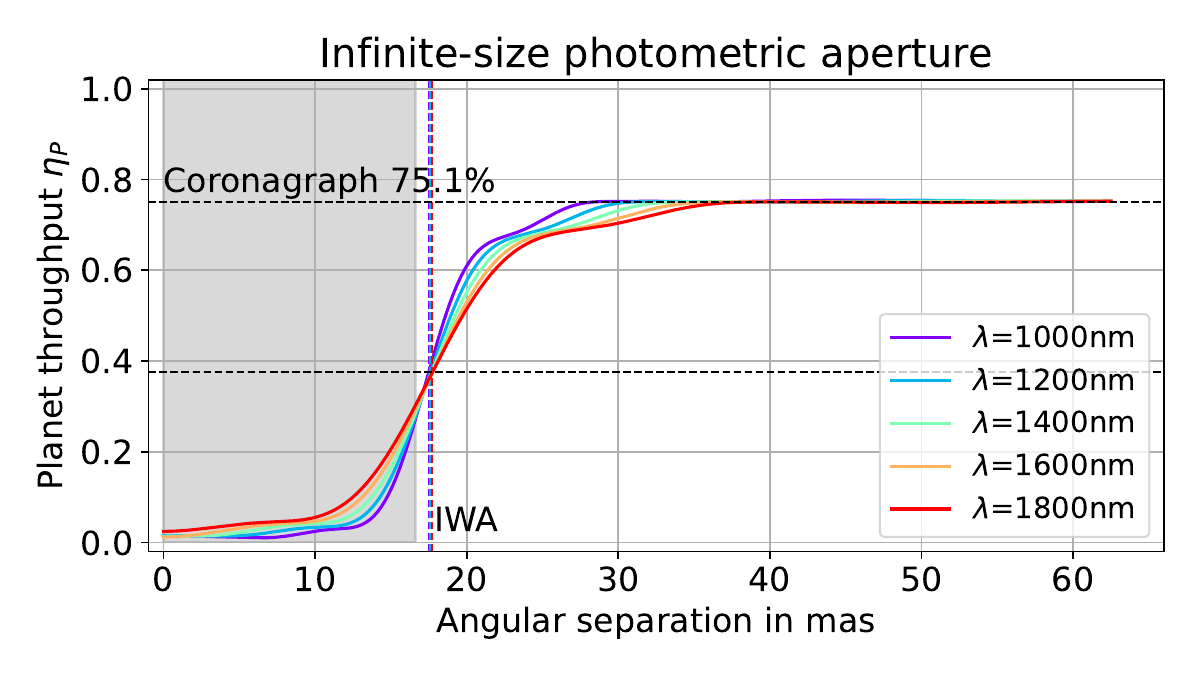}
    \caption{Normalized planet throughput as a function of the angular separation from an observed star at different wavelengths. The grey area denotes the projected opaque mask in the re-imaged focal plane of the coronagraph. The term IWA denotes the inner working angle of the coronagraph. Its value is 17.7mas for our design.}
    \label{fig:throughput}
\end{figure}

These results are encouraging. Still, we need to ensure a good preservation of the planet photons through the coronagraph. We here assess the transmitted flux for a planet (also known as planet throughput) as a function of its angular separation from the star at different wavelengths, see Figure \ref{fig:throughput}. The planet flux is normalized with respect to the incident flux arriving at the ELT. At least, half of the maximum planet flux through the coronagraph of the planet is transmitted at an angular separation of 17.7mas (inner working angle [IWA] of the coronagraph). This value ensures that more than half of the maximum planet flux is transmitted through the coronagraph at 20\,mas.

In terms of absolute transmission, the coronagraph in its current configuration preserves 75\% of the incoming flux of the planet at the level of the ELT pupil, corresponding to an infinite-size photometric aperture. This value is deemed very good in terms of exoplanet transmission. As a side note, the planet flux nearly achieves 50\% of the total incoming light for a photometric aperture of 7\,mas, corresponding to the smallest available spaxel size.  

\section{Contrast Budget}
We now assess the coronagraph performance in terms of contrast at the wavelengths ranging from 1.0-1.7\,$\mu$m (YJH bands). 

The SCAO subsystem should deliver a contrast better than $3.0 \times 10^{-3}$ baseline ($10^{-3}$ goal) on the AO corrected PSF over the 1.0-1.7\,$\mu$m wavelength range and at an angular separation of 20$\pm$3.5\,mas from an observed star with a magnitude I$\leq$8 and in median seeing conditions. The coronagraph module is assumed to be located between the SCAO dichroic and IFU unit. 

As a reminder, the raw contrast is defined as the radial intensity profile (azimuthal average) of the intensity for an image of an observed unresolved on-axis star. All the images are normalized with respect to the intensity peak of the non-coronagraphic image (i.e., PSF). The raw contrast value is averaged over a radial range of $\pm$3.5\,mas around an angular distance of 20\,mas from the star. 

\subsection{Simulation assumptions}
The coronagraph simulations are performed under median atmospheric conditions with the GMT-like SCAO assumptions. PASSATA and then SPECULA\cite{Rossi2026} are libraries developed by INAF-Arcetri and they provide the expected AO corrected phase screens from the SCAO. They include several effects such as AO atmospheric turbulence residuals, windshake effects, and the static co-phasing errors from the ELT primary mirror M1. The current simulations account for the ELT pupil. 

In this study, we assume the absence of (i) the chromatic dispersion due to the atmosphere, (ii) non common path aberrations between the wavefront sensing path from the SCAO and the science path. Such effects will be investigated in Section \ref{sec:error_budget}.

\subsection{Contrast performance and discussion}
\begin{figure}[!ht]
    \centering
    \includegraphics[width=0.67\linewidth]{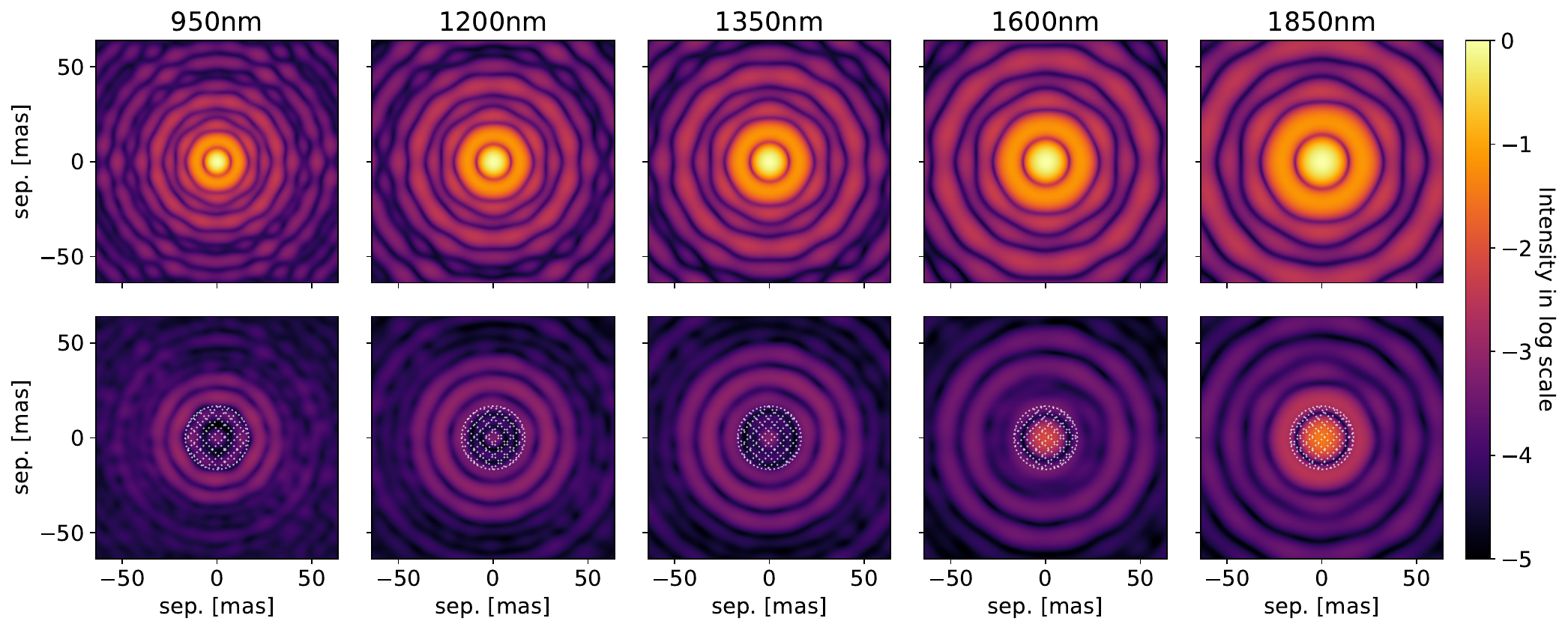}
    \caption{Simulated images of an on-axis point source in median observing conditions after SCAO correction. The top and bottom rows display the frames in log scale of the PSFs and the coronagraphic images with a Lyot coronagraph in YJH configuration, in monochromatic light at 5 different wavelengths ranging from 950 to 1850\,nm. In the bottom row, the dashed disk represents the projected area of the coronagraph focal plane mask in the final image plane.}
    \label{fig:simulated_images}
\end{figure}

Figure \ref{fig:simulated_images} displays the simulated PSF and coronagraphic images in monochromatic light at five different wavelengths ranging between 950 and 1850\,nm in median observing condition (JQM) with the coronagraph configuration in YJH band. At a given wavelength, an attenuation of the intensity level is clearly observed from the AO corrected PSF to the coronagraphic image. The structure of the coronagraphic images reveal a combination of two predominant effects: (i) the speckles due to the atmospheric residuals left after SCAO correction and (ii) the diffraction residuals due to the coronagraph effect. Since the SCAO correction improves toward the red end of the spectrum, the AO residual speckles are prevailing at the shortest wavelengths while the diffraction residuals due to the coronagraph represent the main contributor at the largest wavelengths. 

\begin{figure}[!ht]
    \centering
    \includegraphics[width=0.67\linewidth]{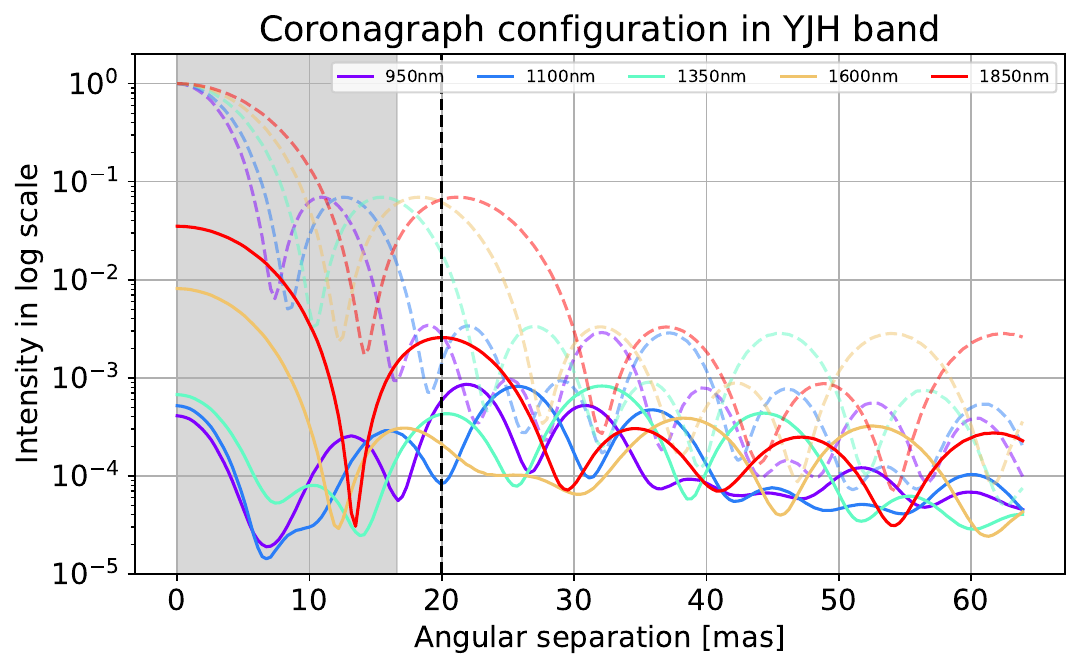}
    \caption{Azimuthally averaged intensity profiles of the images in median observing conditions (JQM) after SCAO correction in monochromatic light at different wavelengths. The dashed and solid lines represent the profiles of the PSFs and the coronagraphic images with a Lyot coronagraph optimized in YJH configuration. The grey area denotes the projected area of the coronagraph focal plane mask in the final image plane. The vertical dashed line denotes an angular separation of 20mas.}
    \label{fig:intensity_profiles}
\end{figure}

This analysis of the images is confirmed by observing the azimuthally averaged intensity profiles of the coronagraphic images at different wavelengths within the considered spectral range and in median seeing conditions, see Figure \ref{fig:intensity_profiles}. The profiles of the coronagraphic images shows an intensity level below $10^{-3}$ at 20\,mas from the star, at all the wavelengths shorter than 1850nm, showing the ability of the Lyot coronagraph to achieve the contrast goal ($10^{-3}$ at 20\,mas) in YJH bands under JQM conditions.

\begin{figure}[!ht]
    \centering
    \includegraphics[width=0.67\linewidth]{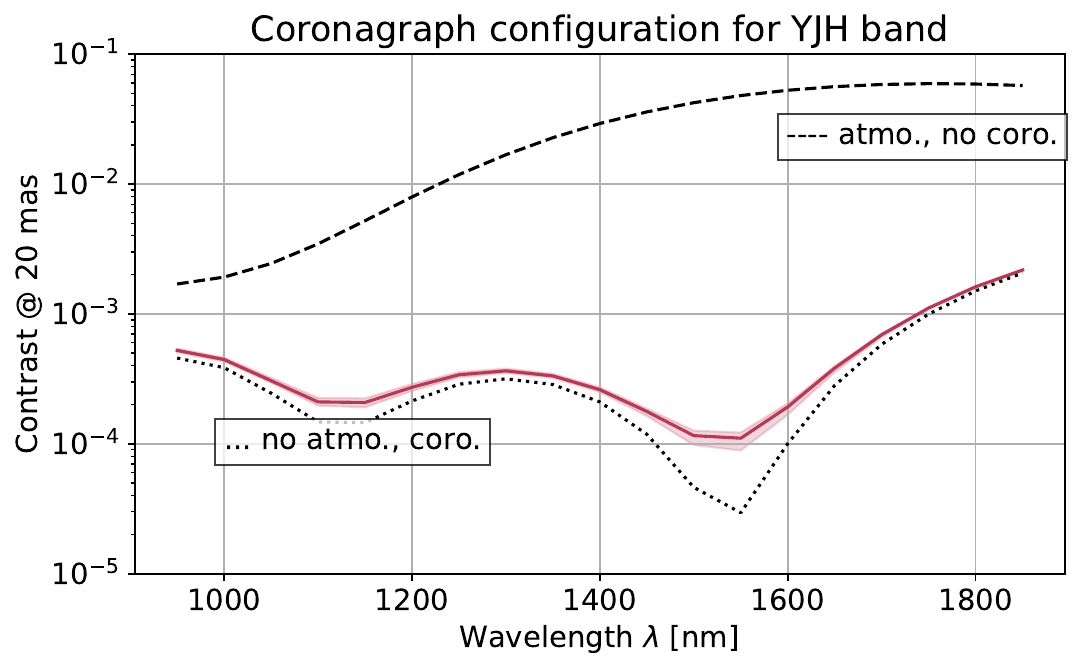}
    \caption{Azimuthally averaged contrast level of the coronagraphic images over an annulus of 7\,mas centered at 20\,mas as a function of the wavelength $\lambda$ in median observing conditions. The plot shows that the image contrast is better than $10^{-3}$ in JQM (median conditions) over 1.0-1.7\,$\mu$m wavelength range. The dashed-line curve represents the contrast with atmospheric residuals and in the absence of the coronagraph. The dot-line curve represents the contrast with no atmospheric residuals and in the presence of the coronagraph. With respect to the non-coronagraphic case in dashed line, the coronagraph provides a contrast gain larger than 10 for wavelengths larger than 1100\,nm in JQM. The envelopes of the curves represent the results for different realizations of AO residuals.}
    \label{fig:contrast_at20mas}
\end{figure}

We extend this study by assessing the contrast in the coronagraphic images at an angular separation 20\,mas as a function of wavelength for different observing conditions, see Figure \ref{fig:contrast_at20mas}. The contrast has been computed over an annular photometric aperture with a width of 7\,mas and centered at an angular distance of 20\,mas from the star. The results confirm the ability of the designed coronagraph in combination with the SCAO to fulfill the contrast requirement. 

\section{Impact of different sources of error}\label{sec:error_budget}

The previous studies do not consider the presence of different effects such as the chromatic dispersion of the atmospheric turbulence, the non-common path aberrations between the wavefront sensing and the science paths in the SCAO-coronagraph combination. In this section, we present a study of these effects to investigate their impact on the coronagraphic performance. In the following, the results are presented with the coronagraph in YJH-band configuration and in median observing conditions (JQM). Each effect is studied at an individual level to establish a preliminary error budget.

\subsection{Pointing error}

In an instrument without coronagraph, errors such as thermo-mechanical drifts or AO tip/tilt residuals introduce a small real-time motion of the position of the star image on the detector. In the presence of a coronagraph, such effects lead to a misalignment of the star image at the intermediate focal plane in which the opaque FPM is located, possibly leading to starlight leaks which impede the observation of planetary companions. Stabilizing the contrast during the observation is therefore crucial to ensure a good image quality and the science return of the instrument. With a coronagraph, a key aspect consists in maintaining the alignment of the star image on the opaque FPM.

We analyze the impact of pointing errors on the coronagraph performance to determine the stability requirements of the star image on the FPM. This study is performed by simulating the injection of tilt errors at the coronagraph entrance pupil plane A to produce the image in the final plane D and estimate the contrast at 20\,mas from the optical axis.

\begin{figure}[!ht]
    \centering
    \includegraphics[width=0.67\linewidth]{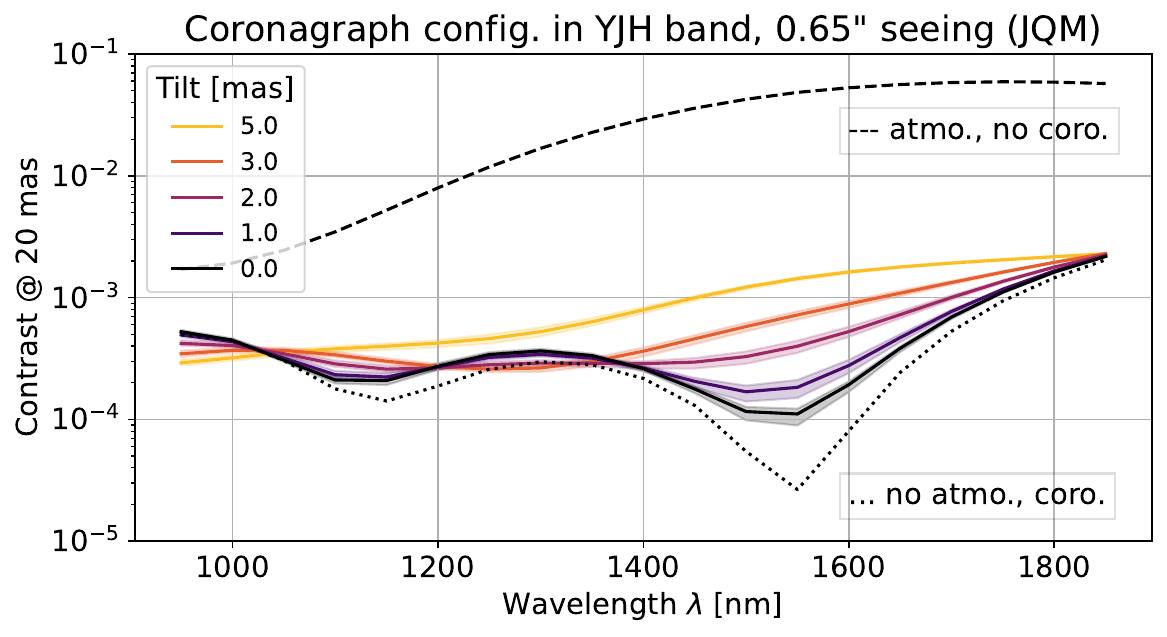}
    \caption{Azimuthally averaged contrast level of the coronagraphic images over an annulus of 7mas width centered at 20\,mas as a function of the wavelength $\lambda$ in median conditions (JQM) for different tilt errors with the coronagraph configuration optimized in YJH band. The dashed-line curve represents the contrast with atmospheric residuals and in the absence of the coronagraph. The dot-line curve represents the contrast with no atmospheric residuals and in the presence of the coronagraph. For the curves with AO residuals and coronagraph, the contrast remains below $10^{-3}$ over the YJH band for pointing errors up to 2\,mas. The envelopes of the curves represent the results for different realizations of AO residuals.}
    \label{fig:pointing_errors}
\end{figure}

Figure \ref{fig:pointing_errors} shows the contrast at 20\,mas at different wavelengths with the coronagraph in the presence of AO residuals for different tilt errors. The contrast provided by the coronagraph remains below $10^{-3}$ over nearly the full YJH bands for tilt errors below 3\,mas. In addition, pointing errors of 2\,mas have no impact on the image contrast. This robustness comes from the use of an opaque FPM with a large diameter (33.4\,mas), making it naturally robust to small pointing errors.

To account for further source errors, we set a preliminary requirement of 2\,mas for residual pointing errors at the level of the coronagraph FPM.

\subsection{Defocus error}
In the previous section, we have investigated the impact on the coronagraph performance of pointing errors which appear in the presence of an offset between the source image and the FPM. The misalignments of the FPM with respect to the optical axis leads to defocus, another low-order aberration mode that can degrade the coronagraph performance.

We analyze the impact of defocus on the coronagraph performance to determine the alignment requirements of FPM along the optical axis. This study is performed by simulating the injection of defocus errors at the coronagraph entrance pupil plane A to produce the image in the final plane D and estimate the contrast at 20mas from the optical axis.

\begin{figure}[!ht]
    \centering
    \includegraphics[width=0.67\linewidth]{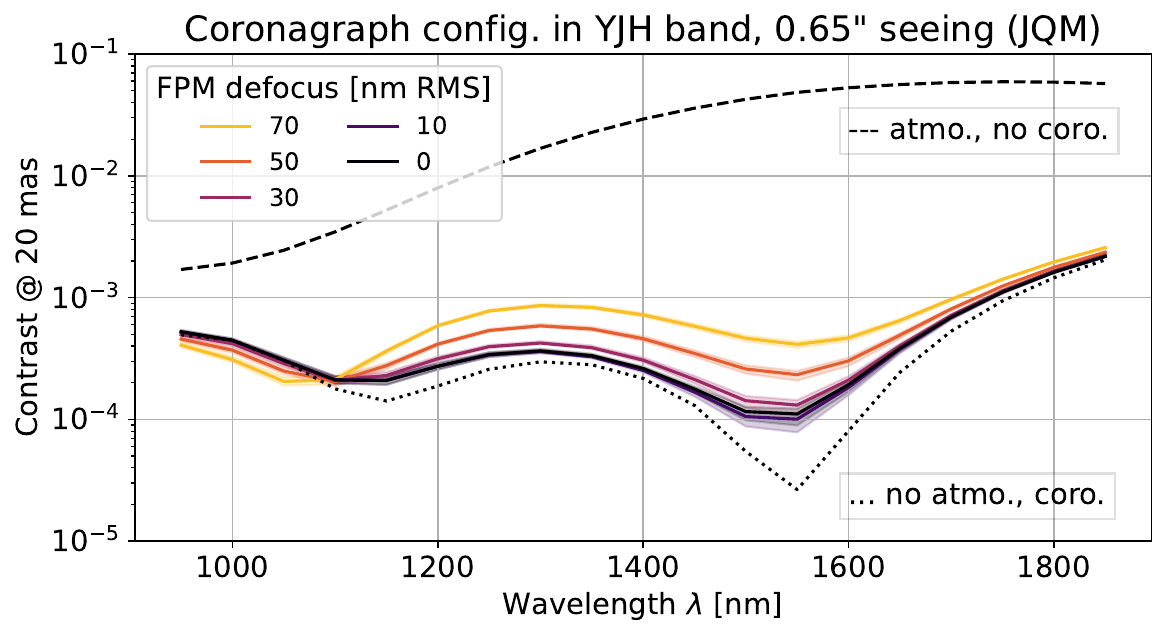}
    \caption{Azimuthally averaged contrast level of the coronagraphic images over an annulus of 7\,mas width centered at 20\,mas as a function of the wavelength $\lambda$ in median conditions (JQM) for different defocus errors with the coronagraph configuration optimized in YJH band. The dashed-line curve represents the contrast with atmospheric residuals and in the absence of the coronagraph. The dot-line curve represents the contrast with no atmospheric residuals and in the presence of the coronagraph. For the curves with AO residuals and coronagraph, the contrast remains below $10^{-3}$ over the YJH band for pointing errors up to 70\,nm RMS. The envelopes of the curves represent the results for different realizations of AO residuals.}
    \label{fig:defocus}
\end{figure}

Figure \ref{fig:defocus} shows the contrast at 20\,mas at different wavelengths with the coronagraph in the presence of AO residuals for different defocus errors. The contrast provided by the coronagraph remains below $10^{-3}$ over nearly the full YJH band for defocus errors below 70\,nm RMS. In addition, defocus errors of 30\,nm RMS have nearly no impact on the image contrast. This robustness comes from the use of an opaque FPM with a large diameter (33.4\,mas), making it naturally tolerant to small defocus errors.

To account for further source errors, we set a preliminary requirement of 30\,nm RMS for residual defocus errors $a_{defocus}$ at the level of the coronagraph FPM. Assuming a F/34.75 beam ratio, it corresponds to a longitudinal defocus error $z_{FPM}$=2$\times a_{defocus} \times$F$^2$= 2$\times$0.030$\times$34.752=72.5\,$\mu$m.

\subsection{Chromatic dispersion}
The atmosphere introduces chromatic errors in the light path that leads to an image of an observed star with a position that depends on the wavelength. In the coronagraphic system, this effect translates into chromatic pointing errors of the star image with respect to the FPM, possibly leading to a contrast degradation at each wavelength.

These chromatic tip/tilt errors can be partially compensated with an active Atmospheric Dispersion Corrector (ADC). We investigate the possibility to add this device in the coronagraph module to stabilize our $10^{-3}$ contrast at 20\,mas in the presence of chromatic dispersion residuals. In the following, we study the amount of residual chromatic dispersion that can be tolerated by the coronagraph to remain below the contrast specification. This will provide some requirements on the residual chromatic dispersion for the ADC design.

\begin{figure}[!ht]
    \centering
    \includegraphics[width=0.67\linewidth]{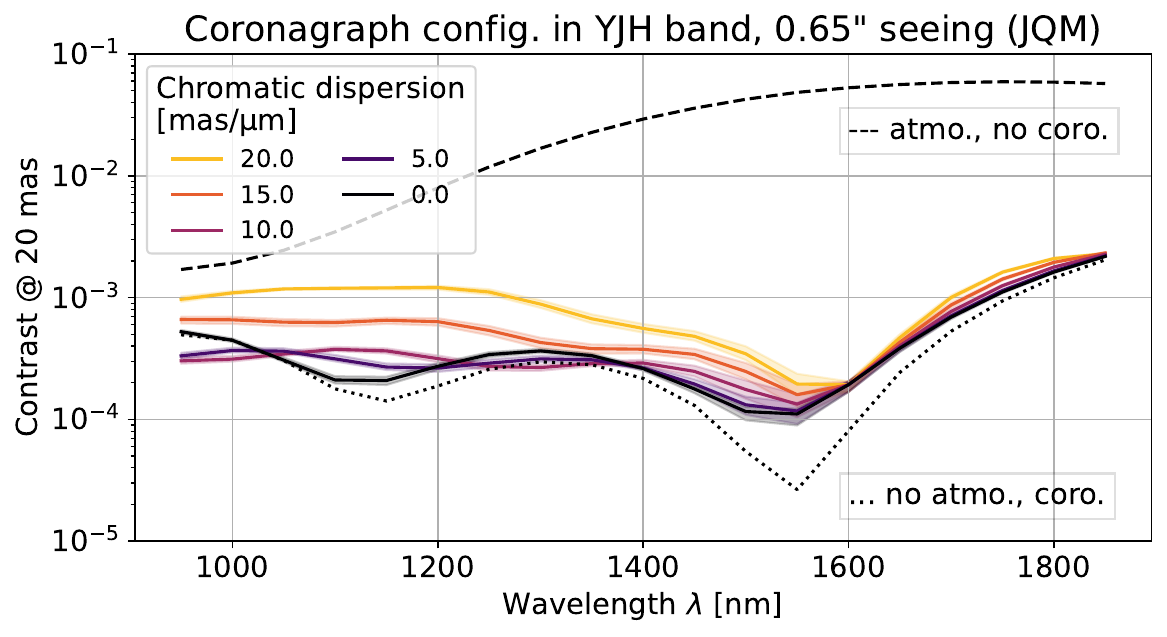}
    \caption{Azimuthally averaged contrast level of the coronagraphic images over an annulus of 7\,mas width centered at 20\,mas as a function of the wavelength $\lambda$ in median conditions (JQM) for different values of chromatic dispersion with the coronagraph. The dashed-line curve represents the contrast with atmospheric residuals and in the absence of the coronagraph. The dot-line curve represents the contrast with no atmospheric residuals and in the presence of the coronagraph. For the curves with AO residuals and coronagraph, the contrast remains below $10^{-3}$ over the YJH band for chromatic dispersions up to 15.0\,mas/$\mu$m. The envelopes of the curves represent the results for different realizations of AO residuals.}
    \label{fig:chromatic_dispersion}
\end{figure}

For our study, we assume a perfect correction of the chromatic dispersion at $\lambda_{ref}$=1.6\,$\mu$m and different residual chromatic dispersions following a linear law with the wavelength. Figure \ref{fig:chromatic_dispersion} shows the contrast at 20\,mas provided by the coronagraph in the presence of AO residuals in median conditions and at different wavelengths as a function of the residual chromatic dispersion (from 0 to 20\,mas/$\mu$m). Our coronagraph is robust to a residual chromatic dispersion up to 15\,mas/$\mu$m with no impact on the expected $10^{-3}$ contrast. Following the study of pointing errors, this tolerance possibly finds its origin in the use of a large opaque FPM for the coronagraph, making the starlight suppression system relatively robust to small pointing errors.

An ADC with a requirement of 5\,mas/$\mu$m for the atmospheric dispersion residuals is reasonably achievable with the appropriate choice of glass for the optics according to the wavelength range of interest.

To account for further source errors and more favorable observing conditions, we therefore set a preliminary requirement of 5\,mas/$\mu$m for the residual atmospheric dispersion for the ADC with the coronagraph module.

\subsection{Lyot stop clocking error}

The coronagraph makes use of a Lyot stop with spider struts which should be aligned with the struts of the ELT pupil to maximize the starlight rejection. Clocking errors can lead to starlight leaks that could deteriorate the contrast in the final image. Specifying the tolerance for the Lyot stop angular position will help us to design the proper opto-mechanical mount and alignment mechanism for this coronagraph part.

We analyze the impact of Lyot stop clocking errors on the contrast at 20\,mas. In our simulations, we propagate the beam through the coronagraph and estimate the contrast for different angular positions of the Lyot stop with respect to its nominal position. 

\begin{figure}[!ht]
    \centering
    \includegraphics[width=0.67\linewidth]{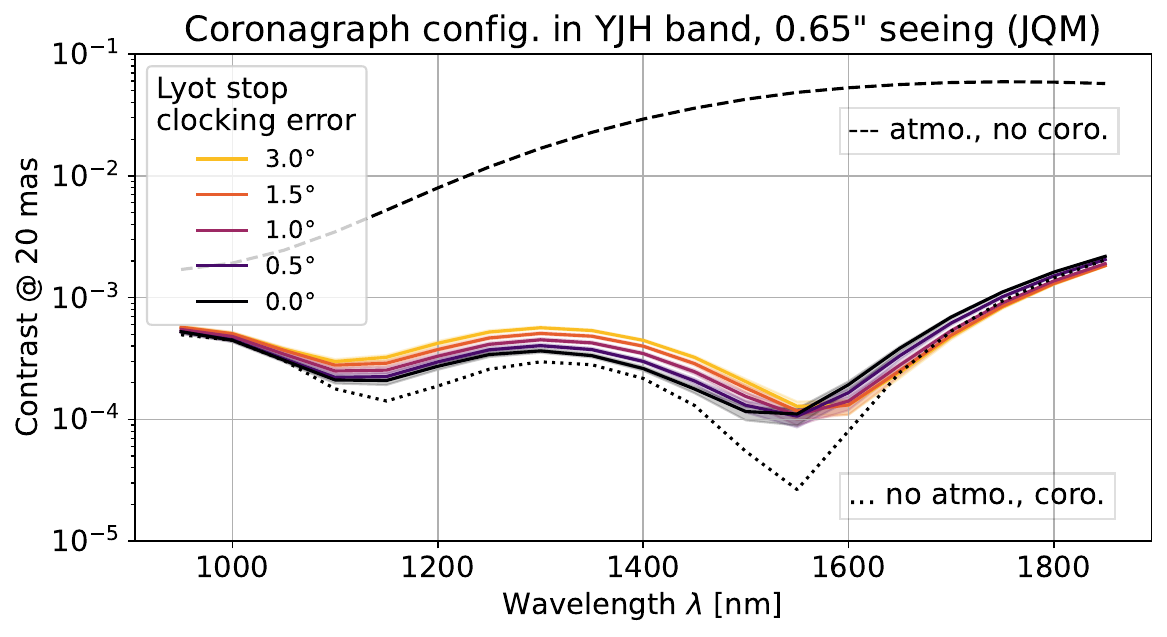}
    \caption{Azimuthally averaged contrast level of the coronagraphic images over an annulus of 7\,mas width centered at 20\,mas as a function of the wavelength $\lambda$ in median conditions (JQM) for different angular positions of the Lyot stop with respect to its optimal position, with the coronagraph configuration optimized in YJH band. The dashed-line curve represents the contrast with atmospheric residuals and in the absence of the coronagraph. The dot-line curve represents the contrast with no atmospheric residuals and in the presence of the coronagraph. For the curves with AO residuals and coronagraph, the contrast remains below $10^{-3}$ over the YJH band for all the considered angular positions of the Lyot stop. The envelopes of the curves represent the results for different realizations of AO residuals.}
    \label{fig:ls_clocking_error}
\end{figure}

Figure \ref{fig:ls_clocking_error} shows the contrast at 20\,mas with this coronagraph at different wavelengths and in the presence of AO residuals in median conditions for different angular positions of the Lyot stop. For an angular position of up to 3$^{\circ}$, the contrast is slightly degraded but largely remains within the specifications. Further tests have been conducted with angular positions errors up to 30$^{\circ}$, showing similar contrast results as the 3$^{\circ}$ angular position error. Our coronagraph is therefore barely impacted by the clocking errors. This result is possibly due to the small amount of residual starlight that is present at the level of the re-imaged ELT struts in the relayed pupil plane C. Their small amount may not significantly impact the contrast with a goal of $10^{-3}$. At deeper contrasts, the impact of the angular position error of the Lyot stop will most likely be more noticeable, as suggested by our contrast curves in our plot with the contrast degradation from 1100 to 1500\,nm when we increase the Lyot stop angular position error.

Based on the existing options of opto-mechanical mechanisms to control the clocking position of the Lyot stop, and to account for further source errors or more favorable observing conditions, we therefore set a preliminary requirement of 1$^{\circ}$ for the angular position of the Lyot stop.

\subsection{Lyot stop lateral positioning error}

Following the Lyot stop clocking errors, the lateral positioning errors of the Lyot stop also lead to stellar leaks that can impact the coronagraph performance. Specifying the tolerance for the Lyot stop lateral position will also be helpful to design the proper opto-mechanical mount and alignment mechanism for this coronagraph part.

We analyze the impact of Lyot stop lateral misalignments on the contrast at 20\,mas. In our simulations, we propagate the beam through the coronagraph and estimate the contrast for different offset positions of the Lyot stop with respect to its nominal position. 

\begin{figure}[!ht]
    \centering
    \includegraphics[width=0.67\linewidth]{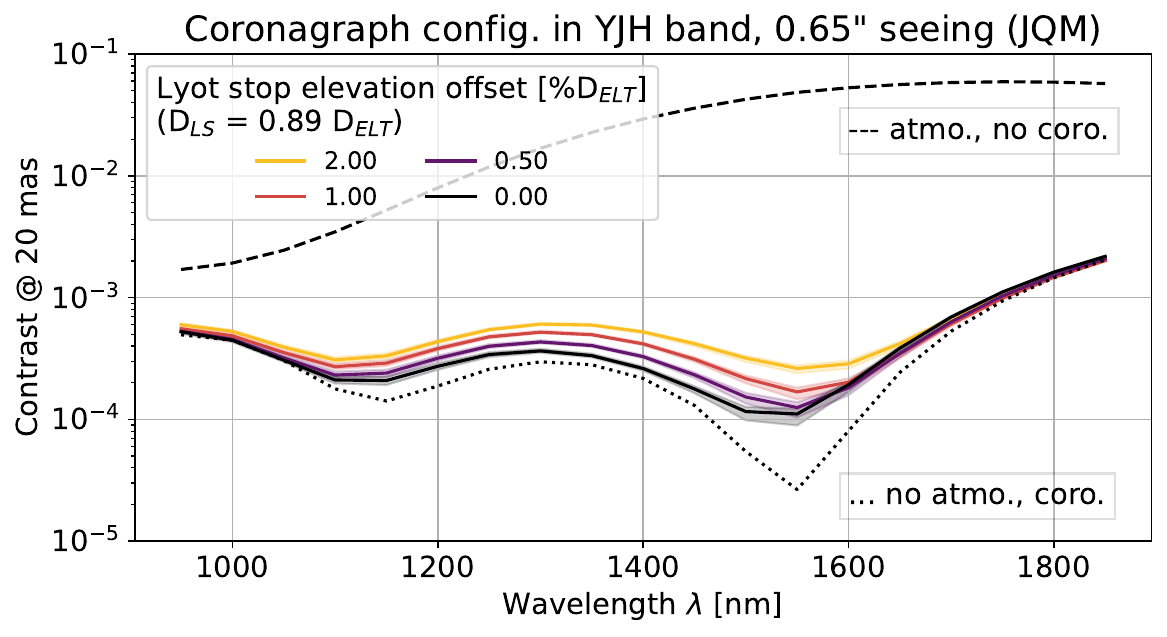}
    \caption{Azimuthally averaged contrast level of the coronagraphic images over an annulus of 7mas width centered at 20mas as a function of the wavelength $\lambda$ in median conditions (JQM) for different vertical offset positions of the Lyot stop with respect to its optimal position, with the coronagraph configuration optimized in YJH band. The dashed-line curve represents the contrast with atmospheric residuals and in the absence of the coronagraph. The dot-line curve represents the contrast with no atmospheric residuals and in the presence of the coronagraph. For the curves with AO residuals and coronagraph, the contrast remains below $10^{-3}$ over the YJH band for all the considered elevation offset positions of the Lyot stop. The envelopes of the curves represent the results for different realizations of AO residuals.}
    \label{fig:ls_lateral_positioning_error}
\end{figure}

Figure \ref{fig:ls_lateral_positioning_error} shows the contrast at 20\,mas with this coronagraph at different wavelengths and in the presence of AO residuals in median conditions for different vertical offset positions of the Lyot stop. For an offset position up to 2\% of the projected ELT pupil diameter $D_{ELT}$, the contrast is slightly degraded but remains within the contrast specifications. Our coronagraph performance is barely impacted by the lateral position error of the Lyot stop with respect to the relayed pupil. This result is most likely due to the small amount of residual starlight that is present outside the pupil in the relayed pupil plane C. Their small amount may not significantly impact the contrast with a goal of $10^{-3}$. At deeper contrasts, the impact of the lateral position error of the Lyot stop will most likely be more noticeable, as suggested by our contrast curves in our plot with the contrast degradation from 1100 to 1500\,nm when we increase the Lyot stop offset position error.

While the present study has been detailed for Lyot stop vertical offsets, simulations with horizontal offsets for the Lyot stop have showed similar contrast results.

Based on the existing options of opto-mechanical mechanisms to control the clocking position of the Lyot stop, and to account for further source errors or more favorable observing conditions, we therefore set a preliminary requirement of 1\% of $D_{ELT}$ for the lateral position of the Lyot stop.

Assuming a projected diameter of the circumscribed ELT pupil of 10.15\,mm, the physical requirement for the lateral position is set to x$_{LS}$=$\pm$101.5\,$\mu$m. 

While we have investigated the impact of the Lyot stop position error at the lateral level, longitudinal error along the optical axis should also be studied. A misalignment of the Lyot stop along the optical axis could lead to Fresnel propagation effects that can alter the coronagraph performance for deep contrasts. A Fresnel propagation model of the coronagraph module will enable to perform simulations and study such effects but it was not available at the time of writing. Such a model will further be investigated with models using HCIpy \cite{Por2018}.

However, the depth of focus at the pupil plane is quite large so the accuracy of positioning of the Lyot stop is not very stringent. Following the study made in SPHERE, we decided to adopt a preliminary value of z$_{LS}$=$\pm$0.5\,mm for the positioning of the Lyot stop. Further studies will enable to determine the position and the stability of the focusing of the Lyot stop. 

\subsection{Non-common path aberrations}

The contrast provided by coronagraph devices can be degraded by different sources of wavefront errors. Among them, these errors can be introduced by the optics and their surface quality. These errors also slowly evolve with time. We here focus on the preliminary specifications of the overall optical surface quality of the optics.

To study this effect, we simulate a phase screen with power spectral density (PSD) following a f$^{-2}$-law, where f denotes the spatial frequency of a given phase ripple. The phase screen is larger than the ELT pupil to allow us to select a random subset with a Brownian motion and simulate errors that evolve with time. This time-evolving error is injected in our coronagraph in the entrance pupil plane A and propagated with AO residuals in median conditions to produce a coronagraphic image and estimate the contrast.

\begin{figure}[!ht]
    \centering
    \includegraphics[width=0.67\linewidth]{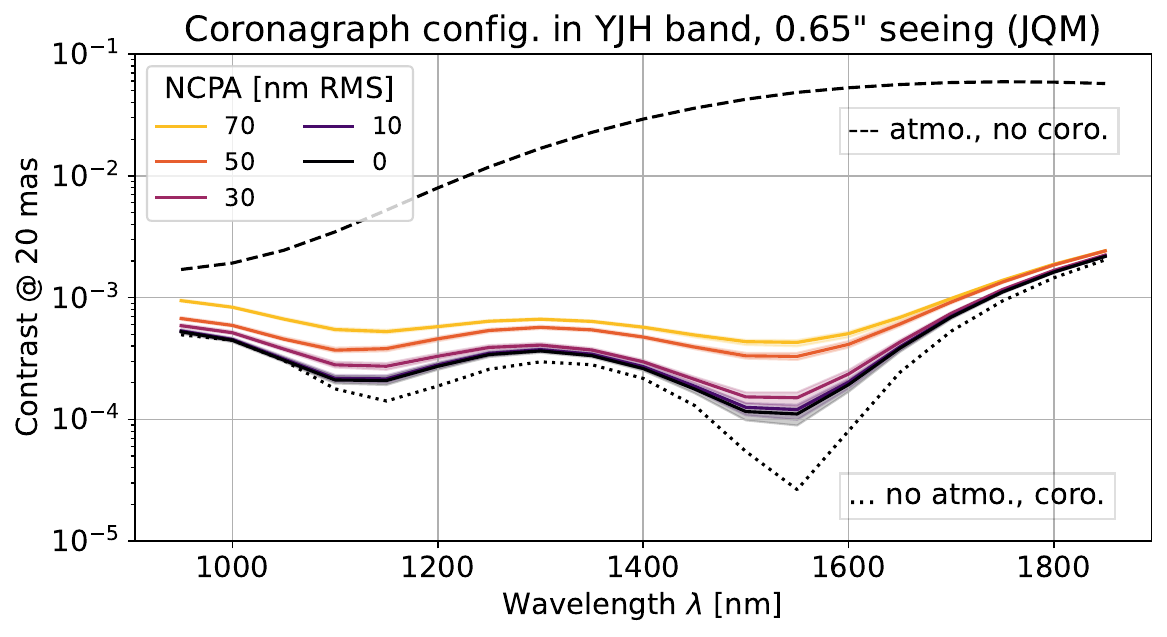}
    \caption{Azimuthally averaged contrast level of the coronagraphic images over an annulus of 7\,mas-width centered at 20\,mas as a function of the wavelength $\lambda$ in median conditions (JQM) for different amounts of optical surface errors. The dashed-line curve represents the contrast with atmospheric residuals and in the absence of the coronagraph. The dot-line curve represents the contrast with no atmospheric residuals and in the presence of the coronagraph. For the curves with AO residuals and coronagraph, the contrast remains below $10^{-3}$ over the YJH band for optical surface errors up to 70\,nm RMS. The envelopes of the curves represent the results for different realizations of AO residuals.}
    \label{fig:ncpa}
\end{figure}

Figure \ref{fig:ncpa} shows the contrast at 20\,mas which is produced by our coronagraph at different wavelengths and with AO residuals in median conditions, as a function of the noncommon path aberrations. The contrast remains below $10^{-3}$ over the YJH band for optical surface errors of 70\,nm RMS. For a given amount of slow-varying optical surface errors, the contrast degradation proves larger at shorter than longer wavelengths since these errors prove larger with respect to the wavelength of interest.

To account for further source errors and more favorable observing conditions, we therefore set a preliminary requirement of 30\,nm RMS for the optical surface errors that are upstream of the coronagraph FPM.

From the 30\,nm RMS acceptable total NCPA, we can determine the amount of aberrations that can be tolerated by the coronagraph at the level of the coronagraph FPM for the low and high-order Zernike aberrations. We use 1000 independent NCPA phase screens with 30\,nm RMS wavefront errors inside the ELT pupil and project each of them into the Zernike polynomial basis following the Noll convention\cite{Noll1976}. We obtain the Zernike coefficients for each Zernike mode for each phase screen and then, average out each Zernike coefficient for a given Zernike mode over the 1000 phase screens. We then RMS sum the coefficients corresponding to the first Zernike modes up to the spherical aberration mode. We finally retrieve the amount for the high order modes by RMS subtracting the total amount with the amount of the low-order modes. This computation gave us:
\begin{itemize}
    \item 11.9 nm RMS for the low-order modes up to the spherical aberrations
    \item 27.5 nm RMS for the high-order modes beyond the spherical aberrations
\end{itemize}

This first iteration gives the amount of wavefront errors that shall be compensated down to 11.9 and 27.5\,nm RMS for the NCPA in low and high order modes at the coronagraph focal plane mask to ensure a $10^{-3}$ contrast at 20\,mas with the coronagraph module. 

\subsection{Summary of our preliminary specifications}
Our preliminary specifications for the error budget with the coronagraph include:
\begin{itemize}
    \item 2 mas pointing errors
    \item 30\,nm RMS defocus error, i.e., a longitudinal defocus z$_{FPM}$=$\pm$72.5\,$\mu$m for F/34.75
    \item 5.0\,mas/$\mu$m chromatic dispersion correction
    \item 1$^{\circ}$ for Lyot stop clocking
    \item 1\% of $D_{ELT}$ for the Lyot stop lateral position, i.e., x$_{LS}$=$\pm$101.5\,$\mu$m for $D_{ELT}$=10.15\,mm, preliminary value of Lyot stop longitudinal position z$_{LS}$=$\pm$0.5\,mm
    \item 30\,nm RMS optical surface errors assuming a f$^{-2}$-power law PSD, which can be decomposed into 11.9 and 27.5\,nm RMS for the low and high order modes. 
\end{itemize}

These values have been used to guide the opto-mechanical design of the coronagraph module\cite{berio2026}.
 
\section{High-contrast capabilities for exoplanet observations}
We have showed that the SCAO assisted with the coronagraph module has been designed to provide a $10^{-3}$ contrast at 20\,mas from an observed star. Based on this estimate, there is a need to assess the ability of ANDES in the SCAO-IFU mode to observe exoplanets in YJH bands. For this purpose, Simonnin  et al. (in prep) have developed APU, an end-to-end simulator to predict the performances of the ANDES in SCAO-IFU mode and determine the detection limits in contrast for different types of exoplanets and the study of circumstellar environments. This simulator uses wavefront errors residuals left from the SCAO subsystem to produce a data cube of images with or without coronagraph for an observed star with a planetary companion at the spatial and spectral resolution (R=100,000) of the IFU subsystem in the 0.9-1.8\,$\mu$m spectral range. APU then uses post-processing techniques to extract the signal of the planetary companion and characterize its physical and chemical parameters.

APU aims to estimate the detection limits in contrast with ANDES and investigate different science cases with this instrument. As a first science case, APU will assess the characterization of gas giant planets in emitted light, studying the atmospheric dynamics, the molecule abundances, the carbon-to-monoxide ratio, some isotope ratios, and others to study the formation and evolution of these planetary companions. In addition, APU will evaluate the capability of the instrument to observe Earth-like planets in reflected light and detect some biosignatures in their atmosphere, in particular for the planets identified as part of the ANDES golden sample\cite{Palle2025}. Finally, APU will explore other science cases such as protoplanetary disks, stellar environments, temperate gas giant planets in reflected light, sub-Neptunes and super-Earths, among many other science cases\cite{Palle2025}.

As a first test case, Simonnin et al. (in prep) uses APU to predict the detection limits of gaseous giant planets with ANDES. In their preliminary study, they investigate the case of L-type planetary objects with an effective temperature of 1600\,K. APU uses BT-Settl \cite{Allard2014} and some homemade models based on the ATMO models \cite{Phillips2020} scaled to high-resolution spectroscopy with PetitRADTRANS \cite{Molliere2019} to simulate the stellar and planetary spectra in the astrophysical scene. The other assumptions include 2h of observations, SCAO residuals in median observing conditions which are simulated using the SPECULA tool \cite{Rossi2026}, telluric transmission from Telfit \cite{Gullikson2014}, emission from SkyCalc \cite{Leschinski2021}, readout noise and dark current from the ANDES Exposure Time Calculator \cite{Sanna2024}, and throughput of the full instrument based on the ANDES requirements. 

\begin{figure}
    \centering
    \includegraphics[width=0.67\linewidth]{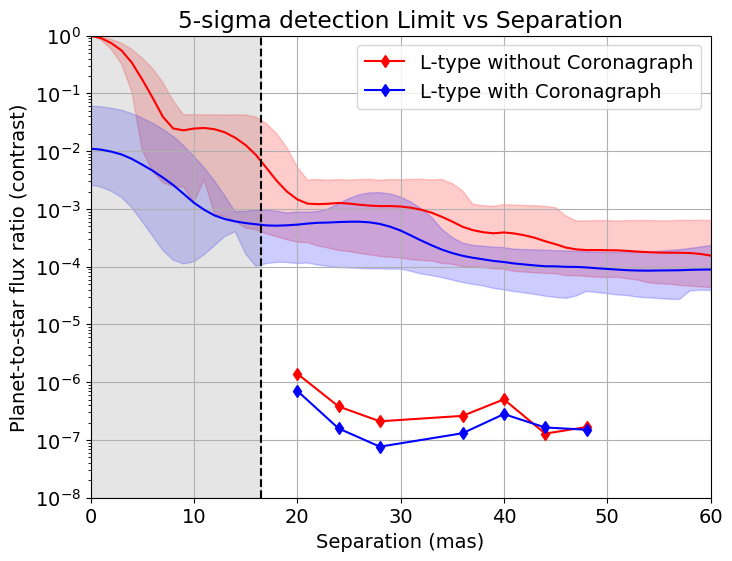}
    \caption{5$\sigma$ detection limits with ANDES using APU for L-type gas giant planets as a function of the angular separation. The top and bottom curves represent the spectrally averaged raw contrast curves and the contrast curves after post-processing using molecular mapping techniques. The red and blue curves represent the curves of the images without and with coronagraph. The envelope of the top curves represents the range of the contrast curves for the wavelengths ranging between 0.9 to 1.8\,$\mu$m. The combination of high-contrast imaging (SCAO+coronagraph) and high-dispersion spectroscopy proves promising with a $10^{-7}$ contrast limit achieved for a L-type companion.}
    \label{fig:5sigma_detection_limits}
\end{figure}

Figure \ref{fig:5sigma_detection_limits} shows the 5$\sigma$ detection limit with ANDES for gas giant planets. The top curves show the spectrally averaged raw contrast curves within the considered spectral band before and after the use of a coronagraph, highlighting some contrast gain at separations larger than 20\,mas by adding the coronagraph. The bottom curves show the contrast curves without and with coronagraph after post-processing via high-dispersion spectroscopy using cross-correlation techniques. The results show detection limits down to $10^{-7}$ for gas giant planets. While these results are preliminary, they prove very promising for the observation of these planetary companions with ANDES. Future work will include more instrumental effects in the simulations to increase the realism and the robustness of the results.

\section{Conclusion and prospects}
In this contribution, we have presented the design and the performance of the CORO module for ANDES in SCAO-IFU mode. A first error budget has been derived to determine the constraints to ensure a $10^{-3}$ raw contrast at 20\,mas from an observed star in median seeing conditions with the SCAO and IFU. The opto-mechanical design is presented in a companion paper \cite{berio2026}, showing a simple and compact design to provide the expected contrast. The development of the CORO module is led by France (Lagrange, LUPM, IRAP, IPAG, LAM) in collaboration with the SCAO group. The preliminary design review of ANDES is planned in October-November 2026 and will include the presence of the coronagraph module in the instrument design.

In the future, the preliminary error budget will be refined with end-to-end simulations to further inform the opto-mechanical design and select an optimal solution for the calibration of the NCPA. Once the design is completed and assembled, the CORO module will be integrated with the SCAO subsystem in Arcetri. In parallel and after a code release expected by the end of 2026, the APU simulator will pursue its development to investigate different exoplanet science cases with ANDES. Such studies will pave the way for the detection and characterization of exoplanetary atmospheres with high-dispersion coronagraphy at the ELT in the 2030s.



\acknowledgments
This work was supported by the Action Spécifique Haute Résolution Angulaire (ASHRA) of CNRS/INSU co-funded by CNES. This work was also supported by the Observatoire de la Côte d’Azur, Région Sud, CNRS and the MERAC Foundation through the 2024 grant awarded to Julia Seidel. The authors gratefully acknowledge their financial and institutional support. 
\bibliography{2026-06_mndiaye_biblio} 
\bibliographystyle{spiebib} 

\end{document}